\documentclass{JINST}
\let\ifpdf\relax

\usepackage{graphicx}
\usepackage{caption}
\usepackage{subcaption}
\usepackage{amsmath}
\usepackage{epstopdf}

\usepackage{appendix}

\newcommand{\nech}{\ensuremath{\mathrm{Ne/(5\%CH_4)}}}

\title{First in-beam studies of a Resistive-Plate WELL gaseous multiplier}

\author{S. Bressler$^a$, L. Moleri$^a$\thanks{Corresponding
author.}, M. Pitt$^a$, S. Kudella$^b$, C. D. R. Azevedo$^c$, F. D. Amaro$^d$, M. R. Jorge$^d$, J. M. F. dos Santos$^d$, J. F. C. A. Veloso$^c$, H. Natal da Luz$^d$, L. Arazi$^a$, E. Olivieri$^e$~and A. Breskin$^a$\\
\llap{$^a$}Department of Particle Physics and Astrophysics, Weizmann Institute of science,\\
76100 Rehovot, Israel\\
\llap{$^b$}Institut f\"{u}r Experimentelle Kernphysik (IEKP), Karlsruhe Institute of Technology,\\P.O. Box 3640 76021 Karlsruhe, Germany\\
\llap{$^c$}I3N, Physics Department, University of Aveiro,\\3810-193 Aveiro, Portugal\\
\llap{$^d$}LIBPhys, Department of Physics, University of Coimbra,\\Rua Larga, PT3004-516 Coimbra, Portugal\\
\llap{$^e$}CERN,\\Meyrin, Switzerland\\

E-mail: \email{luca.moleri@weizmann.ac.il}}

\abstract{We present the results of the first in-beam studies of a medium size (10$\times$10~cm$^2$) Resistive-Plate WELL (RPWELL):a single-sided THGEM coupled to a pad anode through a resistive layer of high bulk resistivity ($\sim$10$^9~\Omega$cm). The 6.2~mm thick (excluding readout electronics) single-stage detector was studied with 150~GeV muons and pions. Signals were recorded from 1$\times$1~cm$^2$ square copper pads with APV25-SRS readout electronics. The single-element detector was operated in \nech  at a gas gain of a few times 10$^4$, reaching 99$\%$ detection efficiency at average pad multiplicity of $\sim$1.2. Operation at particle fluxes up to $\sim$10$^4$~Hz/cm$^2$ resulted in $\sim$23$\%$ gain drop leading to $\sim$5$\%$ efficiency loss. The striking feature was the discharge-free operation, also in intense pion beams. These results pave the way towards robust, efficient large-scale detectors for applications requiring economic solutions at moderate spatial and energy resolutions.}


\begin{document}

\section{Introduction}
\label{sec: Introduction}

The Thick Gas Electron Multiplier (THGEM) was first introduced as a simple, robust detector suitable for applications requiring large detection areas~\cite{Chechick04}. Presently, THGEM-based detectors are being developed and employed in a large variety of basic and applied fields~\cite{Breskin09}. Examples are CsI-coated cascaded-THGEM UV-photon detectors~\cite{Chechick05, Azevedo10} for Ring Imaging Cherenkov (RICH) devices, advantageously replacing wire-chambers~\cite{Alexeev10, Breskin11, Peskov10, Alexeev12, Alexeev14}; cryogenic gas-avalanche detectors~\cite{Buzulutskov12} developed for charge readout in the vapor phase of noble liquid Time Projection Chambers (TPCs) for neutrino physics and rare-event searches~\cite{Resnati11, Bondar12}; cryogenic gaseous photomultipliers (GPM)~\cite{Arazi15} for UV-photon detection in dark matter experiments, medical imaging~\cite{Duval11} and neutron/gamma imaging in cargo inspection systems~\cite{Breskin12, Israelashvili15}; fast-neutron detectors with dedicated converter-foils~\cite{Cortesi12, Cortesi13}; and low-pressure helium-filled detectors for charged-particle track recording in an Active Target TPC~\cite{Cortesi15}. Recent studies have also demonstrated the use of THGEMs immersed in liquid xenon as Liquid Hole Multipliers (LHM)~\cite{Breskin13, Arazi_1_15, Erdal15} developed for recording both scintillation photons and ionization electrons in noble-liquid detectors, with possible applications in future dark matter and neutrino experiments.  Lastly, thin THGEM-based sampling elements are under development by our group for potential application in digital hadronic calorimeters (DHCAL) in future linear-collider experiments~\cite{Arazi12, Arazi13, Bressler13}.
The wide interest in THGEM-based detectors has resulted in the development of production techniques and concepts (for example~\cite{Liu13, Alexeev13}), including the use of resistive films and materials for reducing the effects of occasional discharges~\cite{Bressler_1_13, Peskov07, Charpak09}. In this context, the experience gained with various configurations of THGEM-based DHCAL sampling elements with resistive anodes~\cite{Arazi12, Arazi13_2} has led to the development of a particularly promising candidate - the Resistive-Plate WELL (RPWELL)~\cite{Rubin_RPWELL13}. The RPWELL is a single-sided (i.e. copper-clad on one side only) THGEM electrode, coupled to a segmented readout anode (e.g. pads or strips) through a plate made of material of high bulk resistivity (figure~\ref{fig: RPWELL configuration}). Extensive laboratory studies of the RPWELL~\cite{Rubin_RPWELL13} demonstrated discharge-free operation at high gas-avalanche gains and over a broad ionization range, making the RPWELL concept suitable for the detection of minimum- as well as highly ionizing particles. In~\cite{Rubin_RPWELL13} it was also shown that RPWELL detectors comprising resistive materials with a bulk resistivity of $\sim$10$^9$~$\Omega$cm do not suffer from a significant gain drop - and hence efficiency drop - under high particle flux. In this work we present the first results of RPWELL-detector evaluation with 150 GeV muon and pion beams, conducted at the CERN SPS/H4 RD51 beam-line. Compared to previously tested THGEM-based configurations~\cite{Arazi12, Arazi13, Bressler13}, we show that this very thin (6.2~mm excluding readout electronics) single-stage detector configuration maintains discharge-free operation under muon beams, as well as under high-rate pion beams, while no degradation in the performance is observed. This robust discharge-free behavior in a single-stage - rather than cascaded - configuration, has a major advantage in terms of production costs, thus making the RPWELL highly promising for applications requiring large detection area such as DHCAL, where the instrumented area will be of the order of 4000-7000~m$^2$~\cite{ILC}.

\begin{figure}[h]
\centering
\includegraphics[scale=0.3]{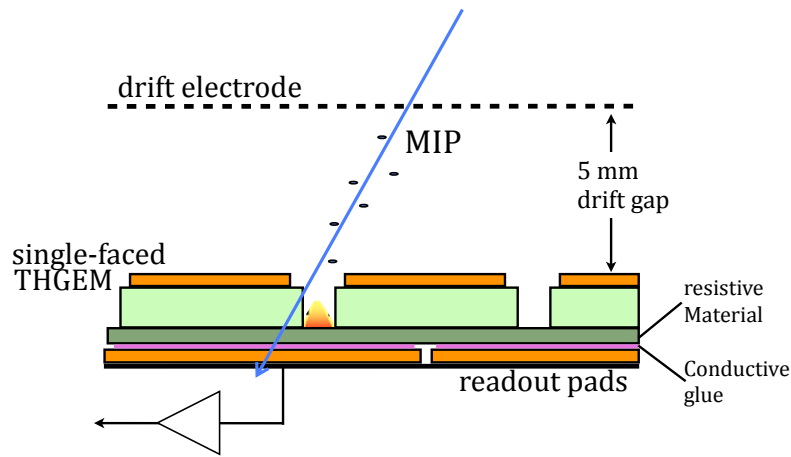}
\caption{The RPWELL configuration. The WELL, a single-sided THGEM, is coupled to a readout anode (e.g. with strips or pads) via a resistive-material plate.}
\end{figure}\label{fig: RPWELL configuration}
\section{Experimental setup and methodology}
\label{sec: Experimental setup and methodology}
\subsection{The RPWELL detector}
\label{sec: RPWELL detector}

\begin{figure}[ht]
\begin{subfigure}{0.5\linewidth}
\caption{}
\includegraphics[scale=0.26]{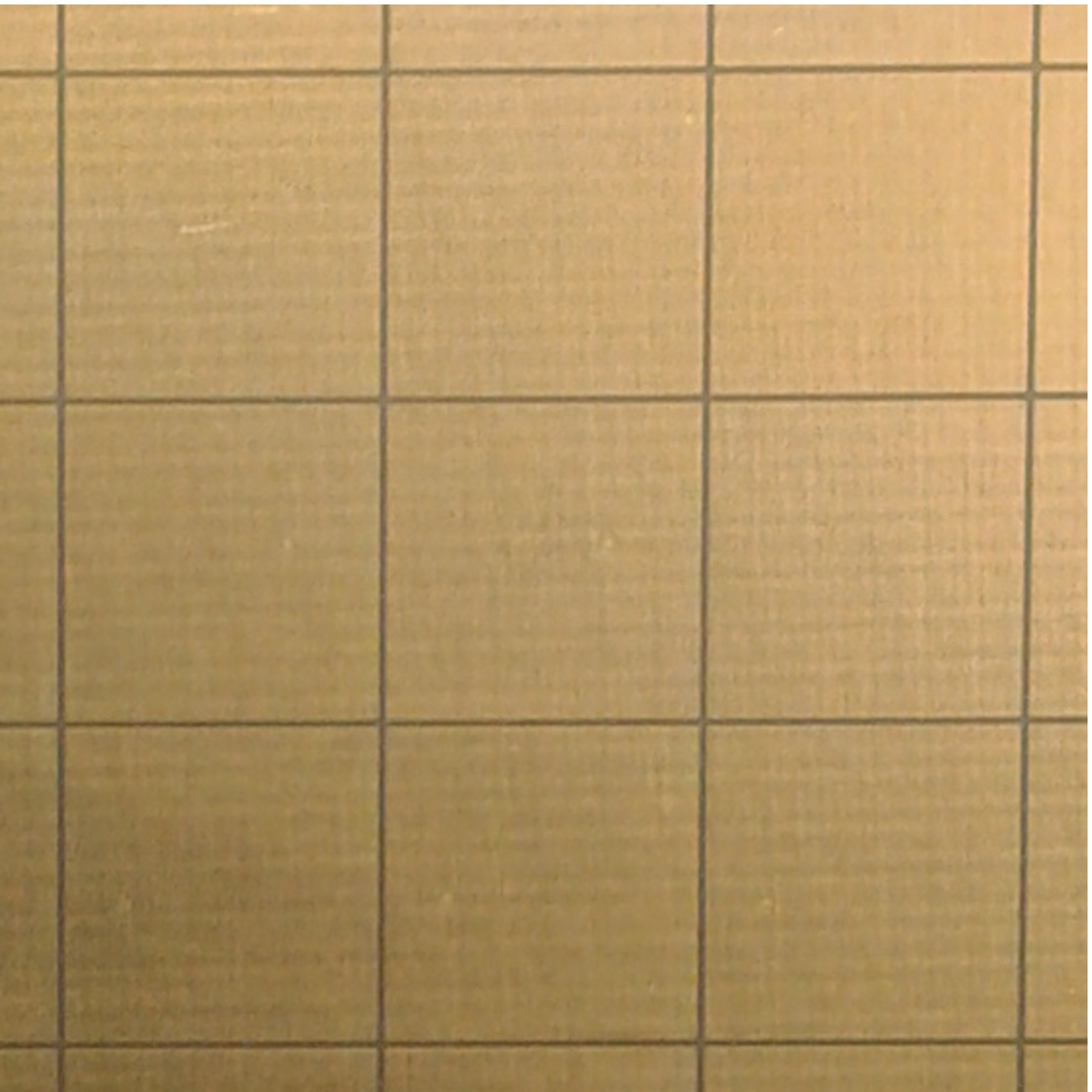}
\end{subfigure}\hfill
\begin{subfigure}{0.5\linewidth}
\caption{}
\includegraphics[scale=0.21]{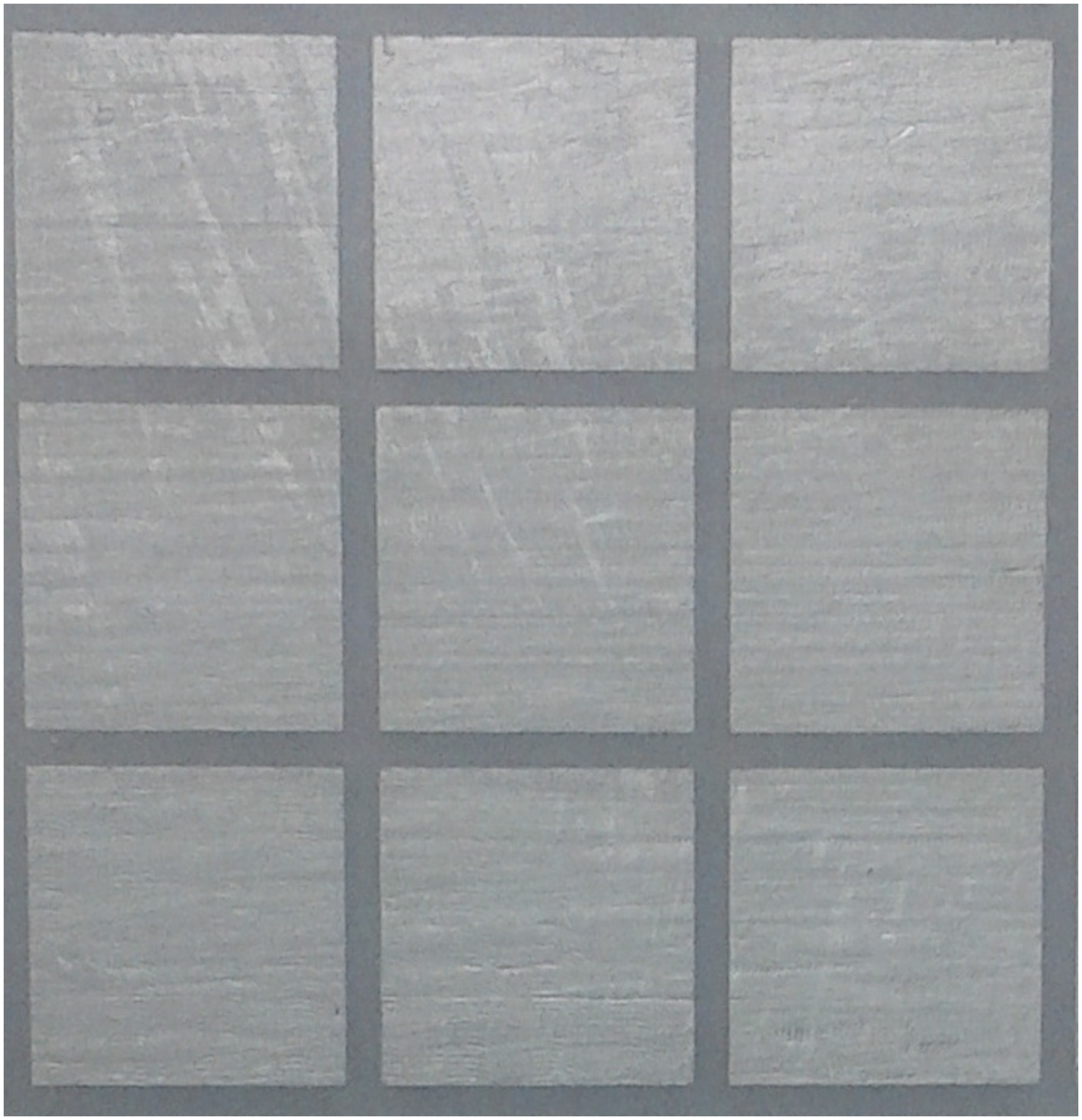}
\end{subfigure}
\caption{Readout 1$\times$1 cm$^2$ anode pads (a) and the matching silver-painted resistive-plate segmentation (b).}\label{fig: anode}
\end{figure}

The THGEM electrode used in this work was 10$\times$10~cm$^2$ in size, manufactured  by 0.5~mm diameter hole-drilling in a 0.8~mm thick FR4 plate, copper-clad on one side only. The holes were arranged in a square lattice with a pitch of 0.96~mm; 0.1~mm wide rims were chemically etched around the holes. The hole-diameter, electrode thickness and rim width were chosen based on previous optimization studies~\cite{Shalem06}. Based on the results in~\cite{Rubin_RPWELL13}, the single-sided electrode was coupled to an 8$\times$8 matrix of 1$\times$1~cm$^2$ pads through a 10$\times$10~cm$^2$ Semitron ESD225\footnote{http://www.quadrantplastics.com} static dissipative plastic plate machined to a thickness of 0.4~mm. The bulk resistivity of this material was measured in~\cite{Rubin_RPWELL13} to be 2$\times$10$^9$~$\Omega$cm.
Good electrical contact between the resistive plate and the readout pads is essential for efficient clearance of the avalanche electrons from the bottom of the resistive material. This was ensured as follows: the bottom side of the plastic was first painted uniformly with conductive silver paint (Demetron Leitsilber 200\footnote{http://www.tedpella.com/technote\textunderscore html/16062\textunderscore TN.pdf}), then, a grid matching the readout pad structure was milled over its surface to avoid charge spread to neighboring pads. Each anode readout pad was glued with conductive epoxy (Circuit Works\footnote{https://www.chemtronics.com/descriptions/document/Cw2400tds.pdf}) to the corresponding painted silver-pad on the resistive plate. Sections of the readout pad electrode and of the matching conductive pads painted on the resistive plate are shown in figure~\ref{fig: anode}-a and ~\ref{fig: anode}-b respectively. 
On the basis of our previous experience with THGEM detector operation in neon mixtures, offering high attainable gas gain at relatively low operation potentials~\cite{Cortesi09, Azevedo10}, the detector was operated in \nech, at a gas flow of 25-50 cc/min. At atmospheric pressure and room temperature, the most probable value (MPV) of the Landau distribution for cosmic muons corresponds to a total number of $\sim$14 ionization electrons along their track in the present 5~mm drift gap; this has been estimated in the laboratory from the ratio between the mean value of the peak of $^{55}$Fe x-rays and the MPV measured for cosmics, and is in accordance with the literature~\cite{Sauli77}. 
The electrodes were biased with individual HV power-supply CAEN A1833P and A1821N boards, remotely controlled with a CAEN SY2527 unit. The voltage and current in each channel were monitored and stored. All HV inputs were connected through low-pass filters. While the RPWELL bias with respect to the grounded anode ($\Delta$V$_{RPWELL}$) was varied throughout the experiment, the drift voltage was kept constant ($\Delta$V$_{drift}$= 250~V corresponding to a drift field of $\sim$0.5~kV/cm).

\subsection{Trigger, tracking and DAQ system}
\label{sec: trigger, tracking and DAQ}

The external trigger system used for event selection is described in~\cite{Karakostas10}; it comprised three 10$\times$10~cm$^2$ scintillators arranged in coincidence. The particle-tracking system, covering a total area of 6$\times$6~cm$^2$, comprised three MICROMEGAS detectors. The RPWELL chamber was placed between two tracker elements. The tracker signals, as well as those recorded from the 64 pads of the RPWELL detector were read out by an SRS readout system ~\cite{Martoiu13} equipped with a single Front-end Concentrator and front-end APV25 hybrids~\cite{Martoiu11}. The SRS data acquisition was triggered by the scintillators coincidence signal. For each triggered event, the amplitude of the analog waveform (sampled in 20 bins of 25~ns each) in each channel was stored; this occurred only for signals crossing a threshold set by a zero-order suppression requirement, as described in section~\ref{sec: analysis framework}. The mmDAQ\footnote{Developed by M.Z.D. Byszewski (marcin.byszewski@cern.ch)} online data acquisition software  was used to store the synchronized data on a PC for further analysis.

The APV25 chip~\cite{APV25}, originally designed for the silicon tracking detectors in CMS, has high
rate capability and low noise. It was operated with a 75~ns shaping time. Due to the low noise level and high sensitivity of the chip, gas gains of the order of few times 10$^3$ are typically sufficient for efficient operation of the detector. As shown in~\cite{Arazi13_2, Rubin_RPWELL13} the rise-time of a typical signal of single-sided THGEM detectors in a WELL configuration is $\sim$1-2~$\mu$s. It comprises a fast component (up to 100~ns in \nech) and a slow one arising from the motion of avalanche ions inside the hole; the latter carries typically about 80$\%$ of the total amplitude. Since the APV25 was operated with 75~ns shaping time, less than 20$\%$ of the total induced signal was collected, requiring the operation of the detector with gas gains of the order of 10$^4$. In what follows we relate to the effective gain, which includes the effect of the short shaping time. This effective gain, estimated from the Landau MPV (section~\ref{sec: RPWELL detector}), is roughly an order of magnitude smaller than the gas gain.

\subsection{Analysis framework}
\label{sec: analysis framework}

The analysis framework, including the noise suppression and the clustering algorithms are described in details in~\cite{Bressler13}. A common zero-order suppression factor (ZSF) sets the threshold of all APV25 channels. For each event the signal in an individual channel is stored only if the following condition is fulfilled:

\begin{equation*}
S > ZSF \cdot (P_{\sigma}\cdot n_{bins})
\end{equation*}

Here S is the sum of the ADC counts in all of the n$_{bins}$ time bins after pedestal subtraction, and P$_{\sigma}$ is the pedestal standard deviation. 

The global detector efficiency was defined as the fraction of tracks matched to a cluster found not more than W~[mm] away from the track trajectory in both x and y directions. The single-event multiplicity was defined as the number of pads in the matched cluster. The average pad multiplicity was defined as the average of the single-event multiplicity. A ZSF of 2 and cluster-track matching parameter W value of 15~mm were found optimal for both muons and pions. We emphasize that in the case of muons a value of W= 8~mm is sufficient for optimal performance. This point, as well as the optimization procedure are described in section~\ref{sec: threshold optimization and cluster-track matching}. 
The detector's discharge probability was defined as the number of discharges divided by the number of hits in the active region of the detector (i.e., in the total area covered by the crossing beam). The number of discharges was extracted directly from the power supply log files by counting the resulting spikes in the current monitor. The pion beam was narrower (a maximum area of 4$\times$3~cm$^2$) than both the acceptance of the scintillators and the tracker detectors. Therefore, the number of pion hits in the active region of the detector was estimated as the number of events where the three scintillators fired in coincidence. Due to the low rate of the muon beam, only pion runs were used to estimate the discharge probability. One should note that pions are prone to induce highly-ionizing secondary events; thus our study yielded an upper limit of MIP-induced discharge probability.

\subsection{Working point: $\Delta$V$_{RPWELL}$, threshold and matching parameter}
\label{sec: threshold optimization and cluster-track matching}

\begin{figure}[ht]
\centering
\includegraphics[scale=0.35]{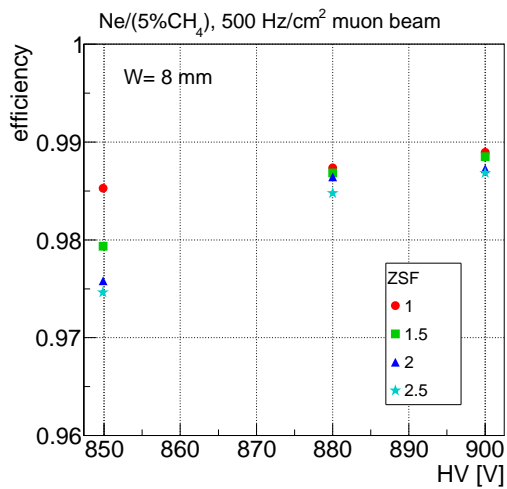}
\caption{The global detector efficiency as a function of the $\Delta$V$_{RPWELL}$ for different ZSF values.}\label{fig: HV scan}
\end{figure}

The working point was adjusted to optimize the detector performance, targeting high detection efficiency at low average pad multiplicity. The latter is a requirement for particle counting, e.g. in a potential application of the RPWELL as a sampling element in Digital Hadron Calorimetry (DHCAL)~\cite{Arazi12}. The optimization was done in three steps using a set of measurements with $\sim$500~Hz/cm$^2$ wide (5$\times$5~cm$^2$) muon beam and a $\sim$13000~Hz/cm$^2$ narrow (2$\times$2~cm$^2$) pion beam. In both beams, only tracks hitting the detector in a 4$\times$4~cm$^2$ area around the beam center were considered. 
\paragraph{Step 1: Choosing the nominal RPWELL operation voltage, $\Delta$V$_{RPWELL}$ - }
Figure~\ref{fig: HV scan} shows the global detector efficiency as a function of $\Delta$V$_{RPWELL}$ for different ZSF values measured in the muon beam. A constant matching parameter W= 8~mm was used. As can be seen in the figure, the global efficiency is already in a plateau at $\Delta$V$_{RPWELL}$= 880~V for all ZSF values applied here. Hence $\Delta$V$_{RPWELL}$= 880~V was selected as the nominal operation voltage. At the incoming particle fluxes used in this measurement, the effective gain (see section~\ref{sec: trigger, tracking and DAQ}) at this operation voltage is $\sim$1700.

\begin{figure}[h]
\centering
\includegraphics[scale=0.35]{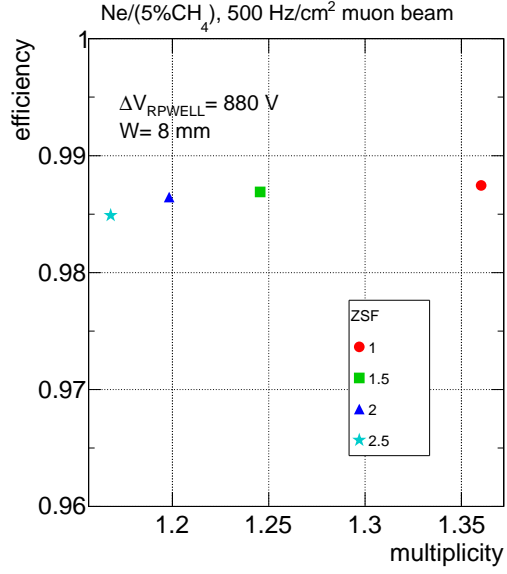}\caption{The global detection efficiency as a function of the average pad multiplicity for different ZSF. The detector bias was kept at the nominal value $\Delta$V$_{RPWELL}$= 880~V.}
\label{fig: ZSF at WP}
\end{figure}

\paragraph{Step 2: Setting the nominal threshold (ZSF) - } Operating at $\Delta$V$_{RPWELL}$= 880~V, figure~\ref{fig: ZSF at WP} shows the global detector efficiency as a function of the average pad multiplicity for different ZSF values measured in a muon beam. Targeting high detection efficiency at low average pad multiplicity, and since similar detection efficiency was recorded with all the ZSF values tested here, the threshold was set at ZSF= 2. Potential bias of the efficiency measurement due to noise hits was estimated in dedicated noise runs. In these runs the data was collected between spills using a clock as a random-like trigger. Processing this runs with a low threshold of ZSF= 0.7, the probability to measure a noise hit in an event was $\sim$10$^{-4}$ which has a negligible effect on the efficiency even with large matching parameter. 

\begin{figure}[ht]
\begin{subfigure}{0.5\linewidth}
\caption{}\includegraphics[scale=0.35]{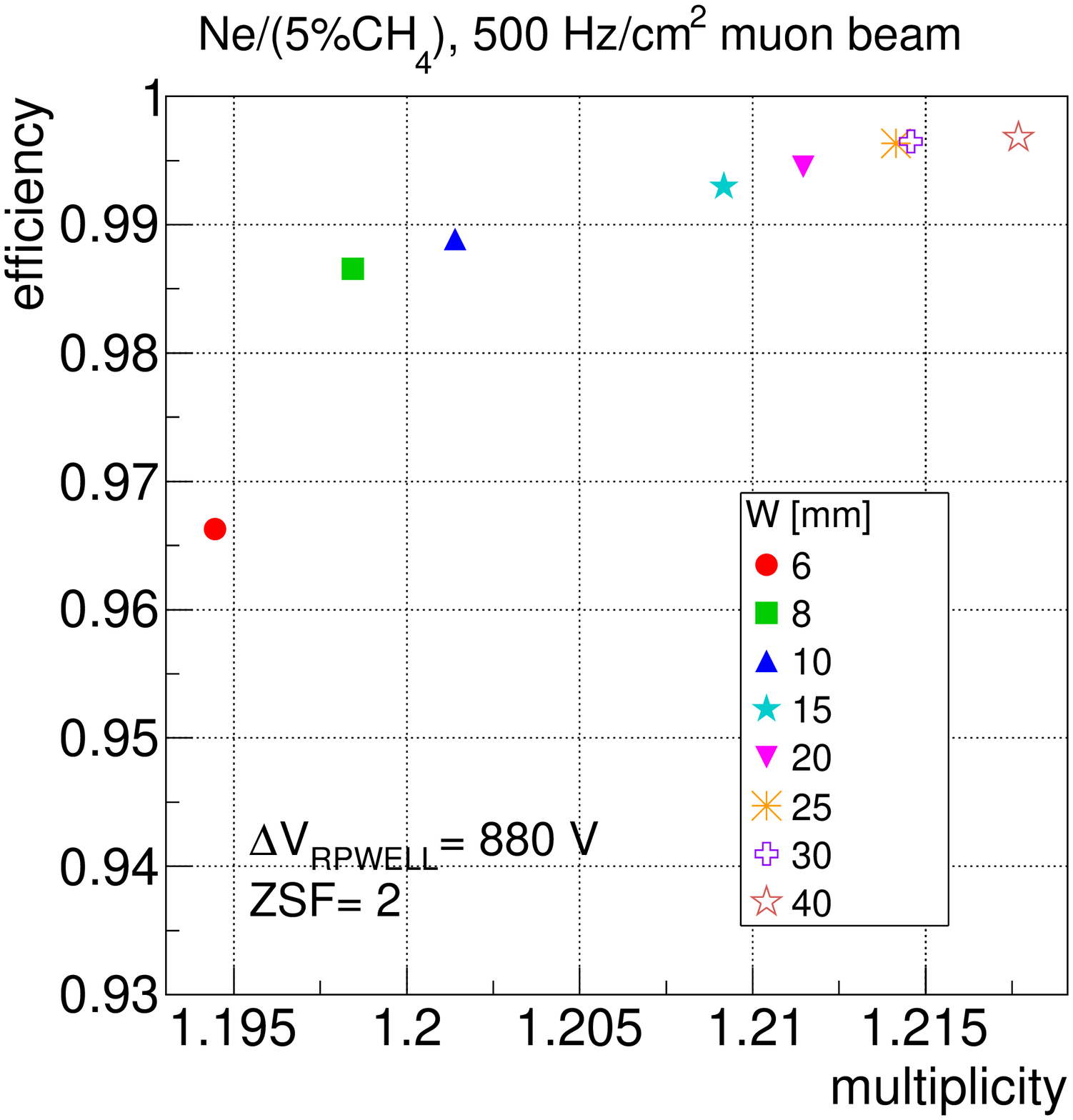}
\end{subfigure}\hfill
\begin{subfigure}{0.5\linewidth}
\caption{}
\includegraphics[scale=0.35]{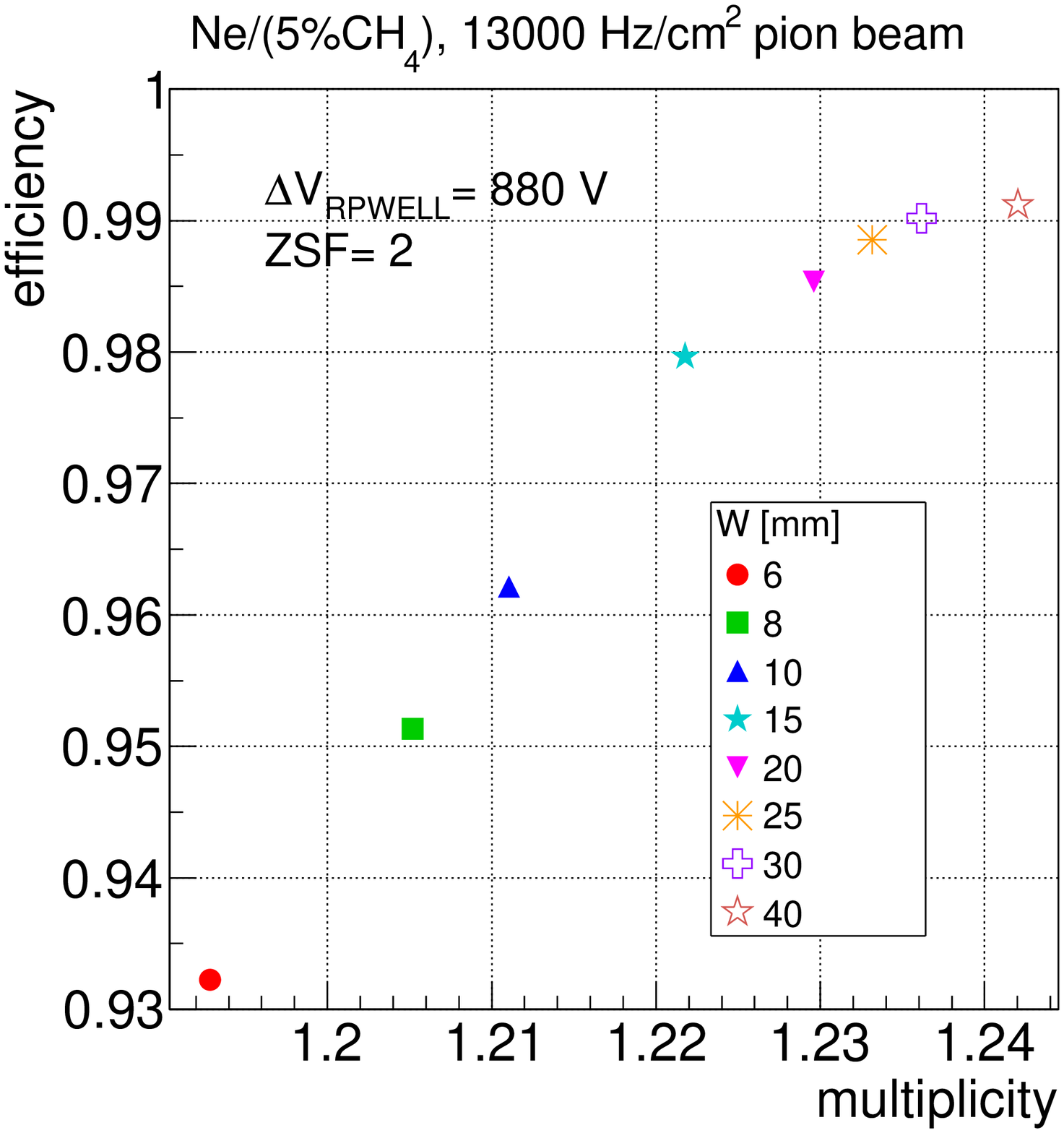}
\end{subfigure}
\caption{The global detection efficiency as a function of the average pad multiplicity for different matching parameter values W. The detector bias was kept at the nominal value $\Delta$V$_{RPWELL}$= 880~V. The analysis was conducted with the nominal thresholds, ZSF= 2 with low rate ($\sim$500~Hz/cm$^2$) muons (a) and high rate ($\sim$13000~Hz/cm$^2$) pions (b). }
\label{fig: W scan}
\end{figure}

\paragraph{Step 3: Fixing the nominal track-cluster matching parameter - } The global detection efficiency as a function of the average pad multiplicity is shown in figures~\ref{fig: W scan}-a and~\ref{fig: W scan}-b for different matching parameter values (W described in~\ref{sec: analysis framework}) as measured in muon and pion beams respectively. The detector was operated at the nominal voltage ($\Delta$V$_{RPWELL}$= 880~V) and the analysis was performed with the nominal threshold (ZSF= 2) chosen above. Using the same matching parameter, lower efficiency is recorded in the pion beam compared to that measured with muons. As can be seen, at larger W values most of this apparent efficiency loss is restored and the detection efficiency reaches a plateau. This indicates that in the pion beam the cluster position is more often found farther away from the track position. This behavior could be explained by secondary particles originating from the incoming pion interactions, affecting the calculated cluster position. This interpretation is also supported by the slightly higher multiplicity recorded with pions compared to that with muons. Targeting high detection efficiency at low average pad multiplicity, the matching parameter was fixed at W= 15~mm for both muons and pions.

\begin{figure}[ht]
\begin{subfigure}{0.5\linewidth}
\caption{}
\includegraphics[scale=0.35]{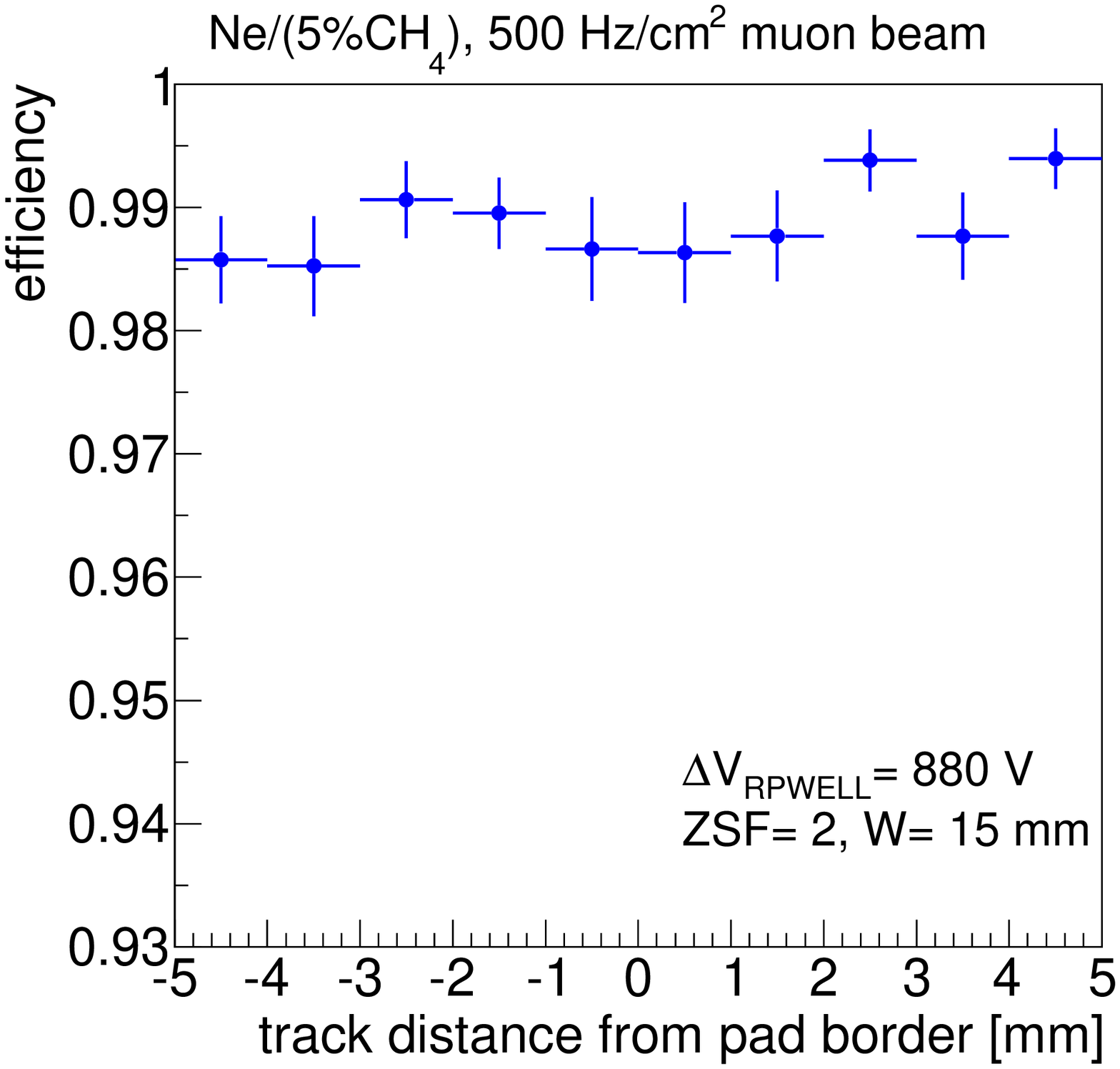}
\end{subfigure}\hfill
\begin{subfigure}{0.5\linewidth}
\caption{}
\includegraphics[scale=0.35]{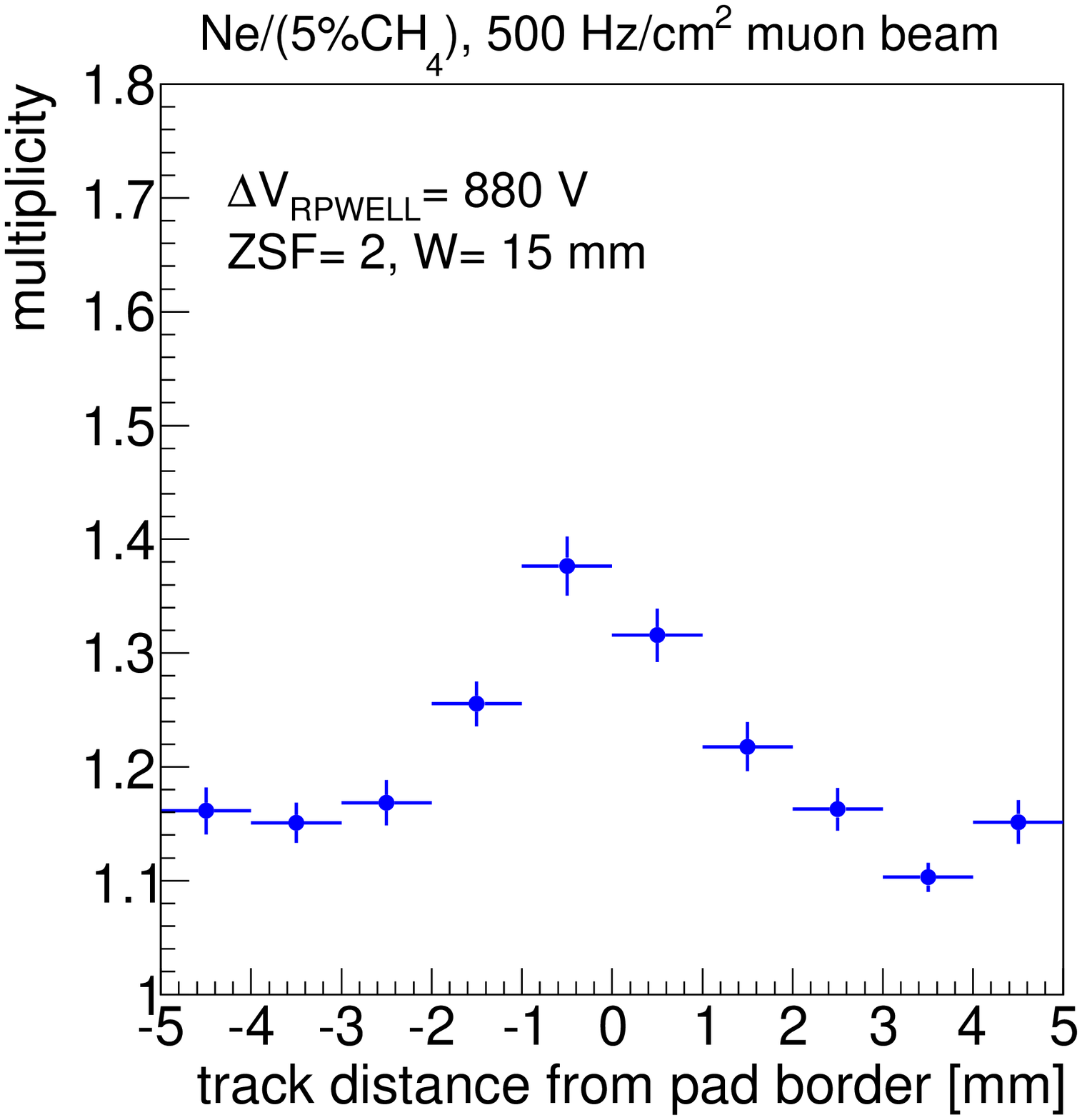}
\end{subfigure}
\caption{The detection efficiency (a) and the average pad multiplicity (b) as a function of the distance from the pad boundaries along one axis on the detector plane. The detector was operated at the nominal conditions under a $\sim$500~Hz/cm$^2$ muon beam. }
\label{fig: local eff and mult}
\end{figure}
\section{Results}
\label{sec: Results}

\subsection{Global and local detection efficiency and average pad multiplicity}
\label{sec: Detection efficiency and pad multiplicity}

Under a low-rate ($\sim$500~Hz/cm$^2$) muon beam and nominal operation conditions, a global detection efficiency greater than 99$\%$ at an average pad multiplicity of 1.21 was recorded (figure~\ref{fig: W scan}-a). Under higher rate ($\sim$13000~Hz/cm$^2$) pion beam and the same nominal operation conditions, the performance of the detector was similar; a global detection efficiency greater than 98$\%$ at an average pad multiplicity of 1.23 (figure~\ref{fig: W scan}-b). 
In figure~\ref{fig: local eff and mult} the detection efficiency (figure~\ref{fig: local eff and mult}-a) and the average pad multiplicity (figure~\ref{fig: local eff and mult}-b) as a function of the distance from the pad boundaries along one axis on the detector plane are shown for operation with $\sim$500~Hz/cm$^2$ muons. The detector was operated at the nominal operation conditions. While the local efficiency is uniform, higher multiplicity is seen to originate from muons traversing the detector close to the pad boundaries resulting in charge sharing between neighboring pads.  

\subsection{Performance under low and high incoming particle fluxes}
\label{sec: rate dependence}

\begin{figure}[ht]
\centering
\includegraphics[scale=0.35]{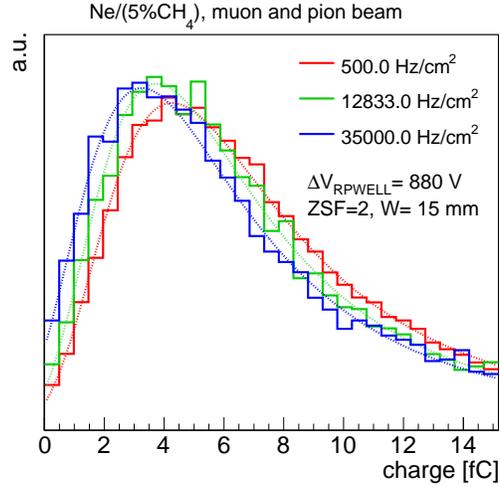}
\caption{Landau spectra measured at three different incoming particle fluxes. The detector bias was kept at the nominal voltage $\Delta$V$_{RPWELL}$= 880~V. The analysis was conducted with the nominal settings, ZSF= 2, W= 15~mm.}
\label{fig: Rate Landau}
\end{figure}

\begin{figure}[ht]
\begin{subfigure}{0.5\linewidth}
\caption{}
\includegraphics[scale=0.35]{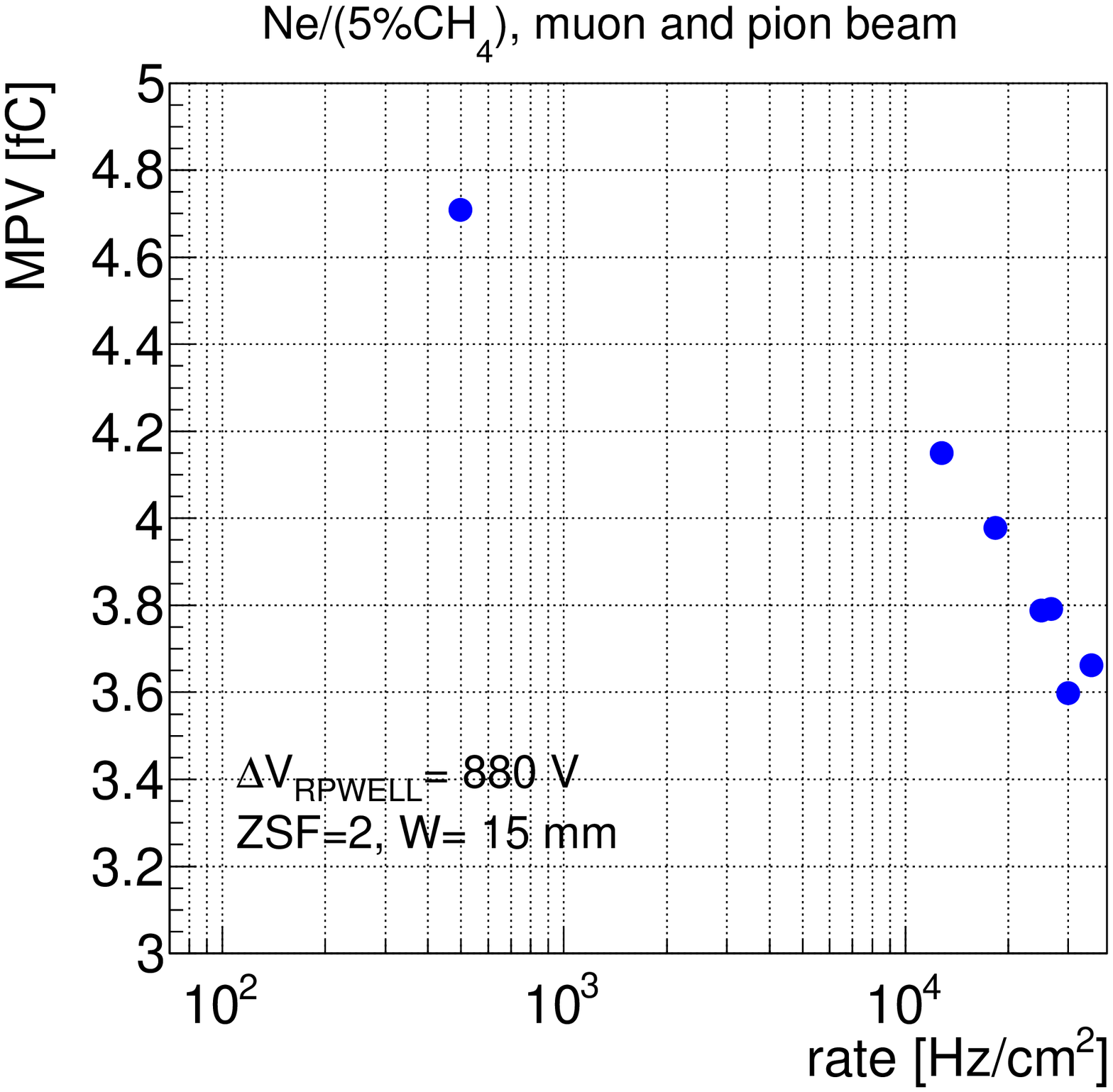}
\end{subfigure}\hfill
\begin{subfigure}{0.5\linewidth}
\caption{}
\includegraphics[scale=0.35]{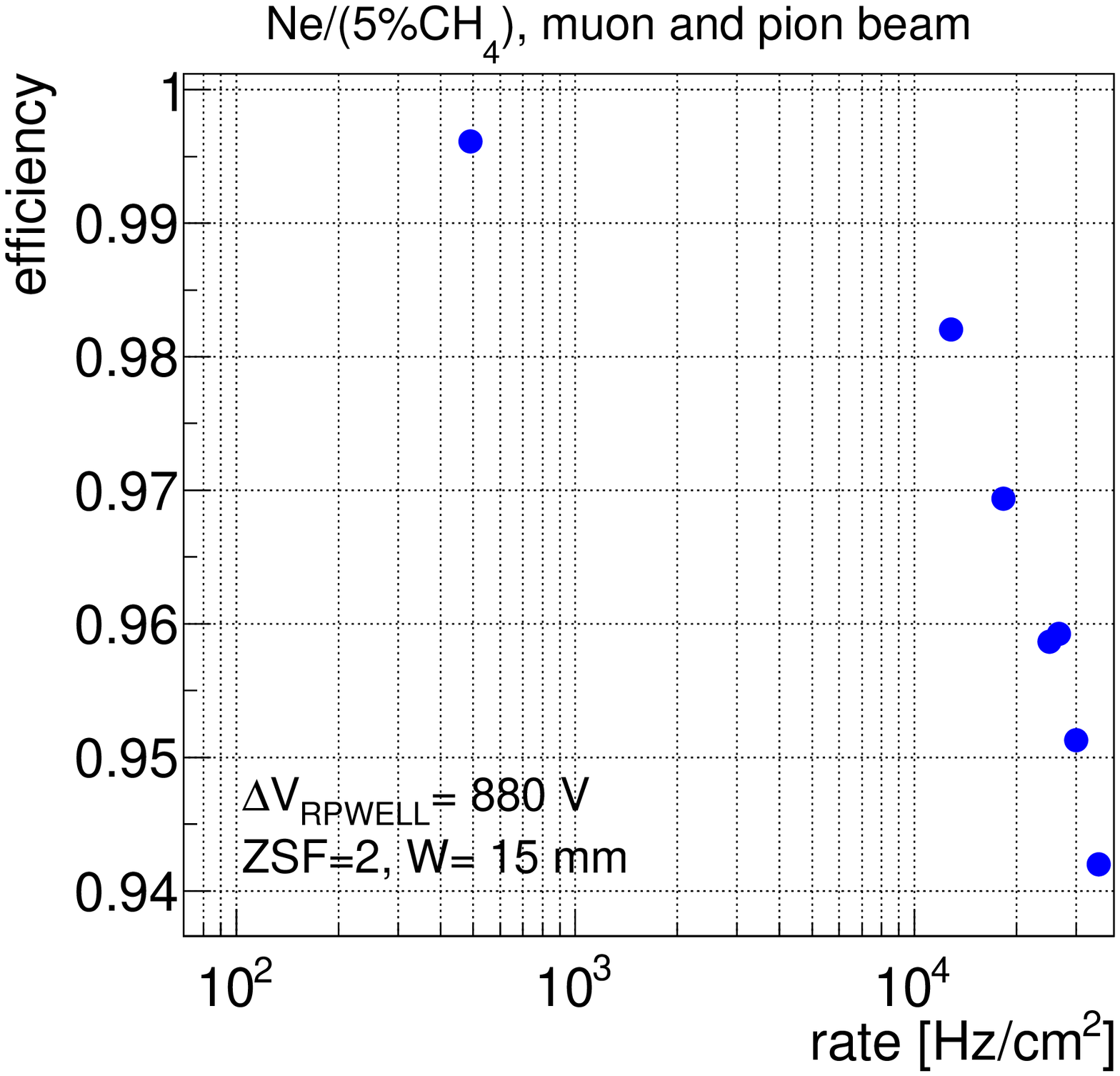}
\end{subfigure}
\caption{Measured charge (a) and global detection efficiency (b) as a function of the incoming particle flux. }
\label{fig: Rate eff MPV}
\end{figure}

In~\cite{Rubin_RPWELL13}, it was shown that the gain dependence on the counting rate was moderate for an RPWELL configuration similar to the one investigated here (0.6~mm thick ESD225 layer, bulk resistivity of $\sim$10$^9$~$\Omega$cm). A gain drop of $\sim$30$\%$ was observed at a counting rate spanning over 4 orders of magnitude.  This study was repeated here with a 5$\times$5~cm$^2$ muon beam, with the lowest flux of $\sim$500~Hz/cm$^2$, and with a narrow (2$\times$3 to 4$\times$3~cm$^2$) pion beam, reaching 35000~Hz/cm$^2$. 
The Landau spectra measured at three different incoming particle fluxes are shown in figure~\ref{fig: Rate Landau}. The detector was operated at the nominal operation voltage, $\Delta$V$_{RPWELL}$= 880~V. The spectra recorded at higher incoming particle flux are shifted towards lower charge values. The measured charge - estimated from the Landau MPV - is shown in figure~\ref{fig: Rate eff MPV}-a as a function of the incoming particle flux. The global detection efficiency as a function of the incoming particle flux is shown, for the same set of measurements, in figure~\ref{fig: Rate eff MPV}-b. A gain drop of $\sim$23$\%$ is noticeable over a counting-rate spanning over more than 2 orders of magnitude; this is in reasonable agreement with the gain drop reported in~\cite{Rubin_RPWELL13} at a similar gain for the same flux of soft x-rays. Over the same range of incoming particle flux, the global detection efficiency dropped by $\sim$5$\%$ from over 99$\%$ to $\sim$95$\%$ by increasing the flux from $\sim$500~Hz/cm$^2$ to 35000~Hz/cm$^2$ (figure~\ref{fig: Rate eff MPV}-b). This drop in the detection efficiency could be avoided by setting the nominal operation voltage slightly higher than 880~V, still maintaining a discharge-free operation as shown in section~\ref{sec: gain stability in time} and~\ref{sec: discharge probability}.

\subsection{Gain stability over time} 
\label{sec: gain stability in time}
 
\begin{figure}[ht]
\begin{subfigure}{0.5\linewidth}
\caption{}
\includegraphics[scale=0.35]{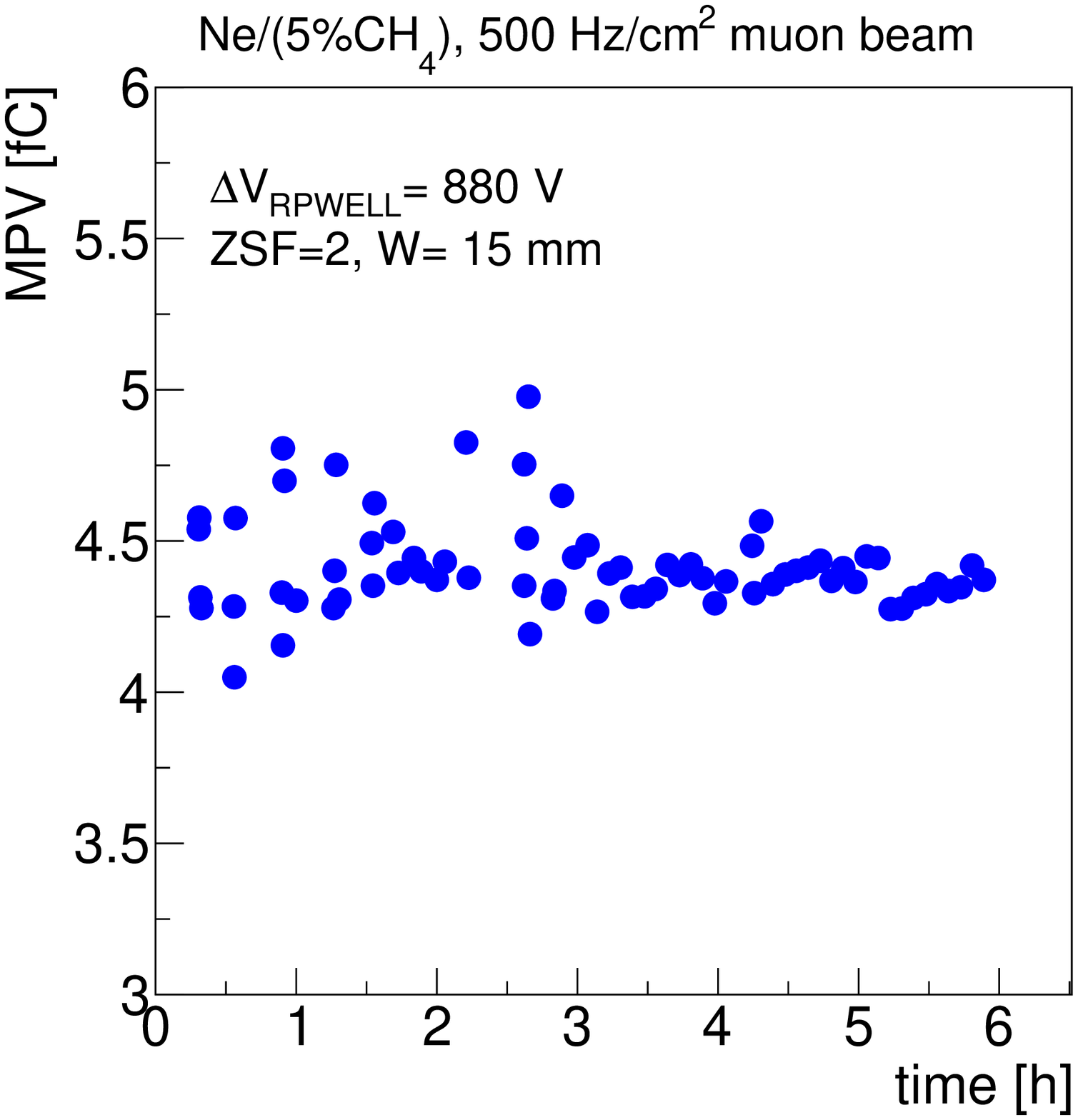}
\end{subfigure}\hfill
\begin{subfigure}{0.5\linewidth}
\caption{}
\includegraphics[scale=0.35]{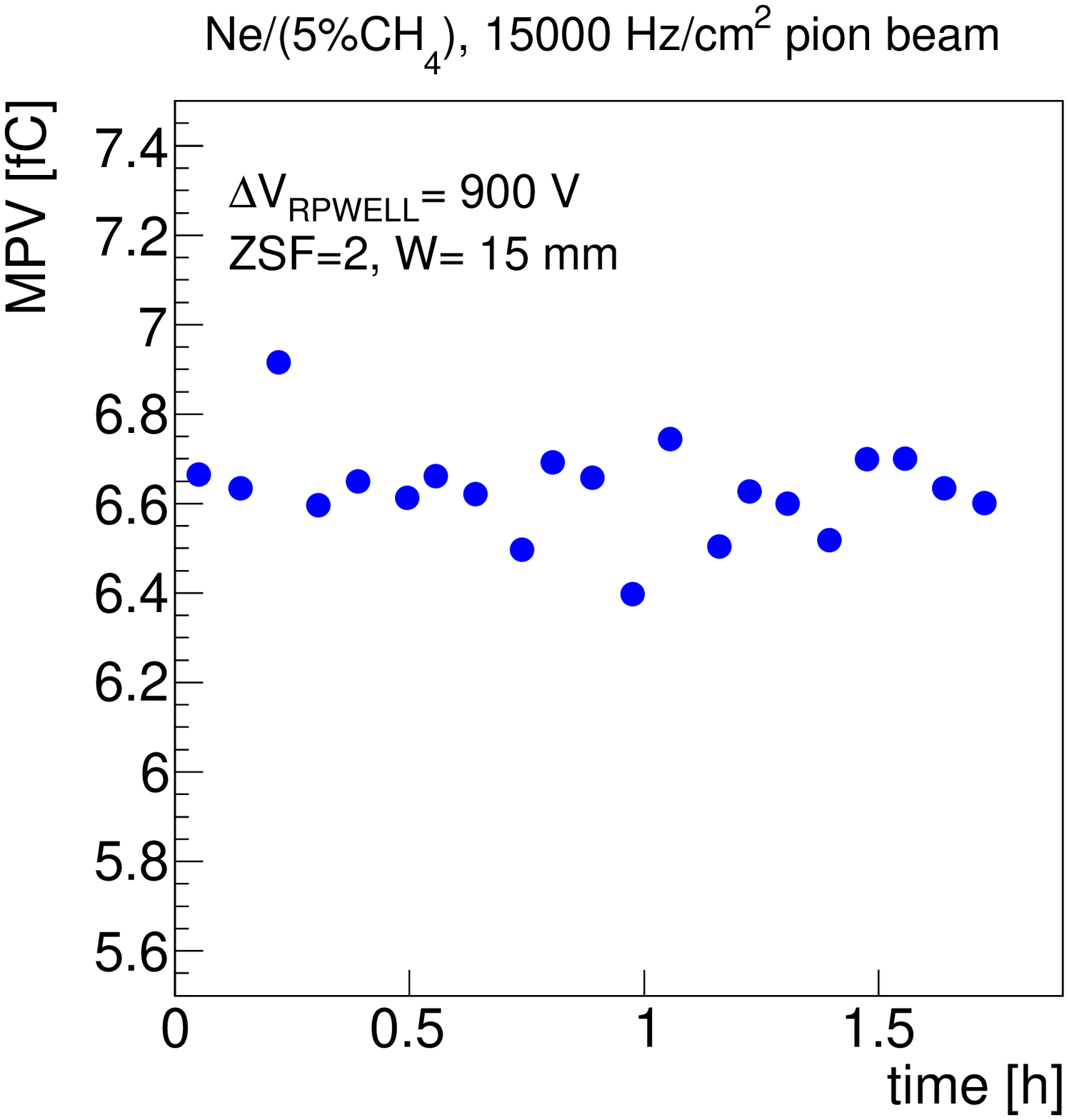}
\end{subfigure}
\caption{Gain stability over time with low rate ($\sim$500~Hz/cm$^2$) muons (a) and high rate ($\sim$15000~Hz/cm$^2$) pions (b).}
\label{fig: Gain stability}
\end{figure}

The detected charge as a function of time is shown in figure~\ref{fig: Gain stability} for measurements carried out with low-rate ($\sim$500~Hz/cm$^2$) muons and high rate ($\sim$15000~Hz/cm$^2$) pions. The detector was operated under the nominal operation voltage with muons while with pions the voltage was increased to $\Delta$V$_{RPWELL}$= 900~V (corresponding to an effective gain of $\sim$3000) to compensate for the slight gain drop with the high-rate pions (section~\ref{sec: rate dependence}). No significant gain variations were observed along the six hours of operation under the muon beam (figure~\ref{fig: Gain stability}-a) (RMS= 0.2~fC), and the two hours of operation under the pion beam (figure~\ref{fig: Gain stability}-b) (RMS= 0.1~fC).  The measurement with muons showed a global detection efficiency of 99$\%$ at an average pad multiplicity of 1.2, while the pion beam showed a global efficiency of  more than 98$\%$ at an average multiplicity of 1.25. As explained earlier, this small efficiency loss could be partially recovered selecting a higher matching parameter W, as also indicated by the slightly higher multiplicity.

\subsection{Discharge probability}
\label{sec: discharge probability}

As mentioned in section~\ref{sec: analysis framework}, a discharge was defined as an abrupt increase in the current supplied to the detector. Due to the low rate of the muon beam, the discharge probability was measured with pions. As an example, during the 1.5 hours of  measurement presented in figure~\ref{fig: Gain stability}-b, no current activity was recorded on any of the detector electrodes. At an incoming particle flux of $\sim$15000~Hz/cm$^2$, this corresponds to discharge-free operation with over 10$^8$ pions resulting in a discharge probability of less than 10$^{-8}$. It is worthwhile mentioning that under these conditions the detector had over 98$\%$ global detection efficiency. One should also note that pions are prone to induce highly-ionizing secondary events; thus this study yields an upper limit of discharge probability with MIPs. Similar behavior was recorded throughout longer periods, supporting the claim that the RPWELL is a discharge-free detector.

\section{Summary and discussion}
\label{sec: Summary and discussion}

The results of a first in-beam study of an RPWELL detector with a Semitron ESD225 resistive plate were presented. This thin, single-stage detector was operated with \nech at variable incoming muon and pion fluxes. 
High detection efficiency (greater than 99$\%$) at low average pad multiplicity ($\sim$1.21) were demonstrated as well as stable operation over time. Discharge-free operation was recorded for the first time with a single-stage THGEM-based detector under pion flux as high as $\sim$15000~Hz/cm$^2$ at a global detection efficiency greater than 98$\%$, maintained throughout the entire measurement. No electrical instabilities were observed over more than 10$^8$ pion events, resulting in recorded discharge probability lower than 10$^{-8}$ in the hadronic beam. 

These results could be compared to previous studies conducted with other THGEM-based detector configurations; single- and double-stage THGEM detectors operated with an induction gap~\cite{Arazi12} and single- and double-stage detectors based on the Segmented Resistive WELL concept~\cite{Arazi13, Bressler13}. The RPWELL detector presents similar or better performance in terms of detection efficiency and pad multiplicity with the major advantage of being completely discharge free.

Compared to other technologies explored under similar conditions, the performance of the 10$\times$10~cm$^2$ RPWELL  detector reported in this work, with respect to detection efficiency and pad multiplicity, is better than that of 1$\times$1~m$^2$ RPCs~\cite{Bilki08, Affatigato15} and 30$\times$30~cm$^2$ GEM detectors~\cite{Yu12} and similar to that of 1$\times$1~m$^2$ MICROMEGAS~\cite{Adloff09, Chefdeville14}.

These results pave the way towards robust, efficient large-scale detectors for applications requiring economic solutions at moderate spatial and energy resolutions. For these purposes, other resistive materials as well as different gas mixtures may be considered.

\acknowledgments

This research was supported in part by the I-CORE Program of the Planning and Budgeting Committee and The Nella and Leon Benoziyo Center for High Energy Physics. A. Breskin is the W.P. Reuther Professor of Research in the Peaceful use of Atomic Energy. F. D. Amaro acknowledges support by FCT under Post-Doctoral Grant SFRH/BPD/74775/2010.

\end{document}